\documentstyle[psfig]{aa}

\newcommand{\nc}{\newcommand}
\nc{\Porb}{$P_{\rm orb}$\,}
\nc{\Teff}{$T_{\rm eff}$\,}
\nc{\logg}{log\,$g$\,}
\nc{\kms}{\,${\rm km\,s}^{-1}$\,}
\nc{\Msun}{$M_{\odot}\ $}
\nc{\Mcz}{$M_{CZ}\ $}
\nc{\vsini}{$v \sin i$ \,}
\nc{\Ali}{A$_{\rm Li}$}

\begin{document}

\title{Observational constraints for Lithium depletion before the
    RGB.\thanks{Based on observations collected at ESO, La Silla, Chile,
    and at the Observatoire de Haute Provence, France, operated by the
Centre National
de la Recherche Scientifique (CNRS)}}

 \author{   P. de Laverny           \inst{1},
            J. D. do Nascimento Jr  \inst{2},
            A. L\`ebre              \inst{3}
      \and  J. R. De Medeiros \inst{2}
  }

    \offprints{J. R. De Medeiros: renan@dfte.ufrn.br}

  \institute{
    Observatoire de la C\^ote d'Azur,
    D\'epartement Fresnel, UMR 6528 CNRS,
    BP 4229, 06304 Nice, France
        \and
    Departamento de F\'{\i}sica,
    Universidade Federal do Rio
    Grande do Norte, 59072-970
    Natal,  RN., Brazil
        \and
    Groupe d'Astrophysique, UMR 5024/CNRS, U. de Montpellier,
    Place Bataillon, 34095 Montpellier, France
      }

  \date{Received 4 February 2003/ Accepted 18 April 2003}

\abstract{ Precise Li abundances are determined for 54 giant stars mostly
evolving across the Hertzsprung gap. We combine these data with rotational
velocity and with information related to the deepening of the convective zone
of the stars to analyse
their link to Li dilution in the referred spectral
region. A sudden decline in Li abundance paralleling the one already
established
in rotation
is quite clear.
Following similar results for other stellar luminosity classes and spectral
regions, there is no linear relation between Li abundance and rotation, in
spite of
the fact that most of the fast rotators present high Li content. The
effects of
convection in driving the Li dilution is also quite clear. Stars with high Li
content are mostly those with an undeveloped convective zone, whereas
stars with a
developed convective zone present clear sign of Li dilution.
}
\maketitle
\markboth{de Laverny, do Nascimento,  L\`ebre,  De Medeiros}{Constraints
  for Lithium depletion before the RGB}

\section{Introduction}

Despite the important advances made in the past decade in the study of the
stellar lithium behavior, a large number of questions are not yet answered.
We do not completely understand the processes controlling lithium production,
nor how and when lithium is depleted.  Related to these
questions are the physical bases of the mixing mechanisms in stellar
interiors. Following the initial study by Bonsack (1959)
different works have attempted to establish the
 behavior of lithium abundance along the giant branch
(e.g.: Alschuler 1975; Wallerstein 1966; Brown et al. 1989) and its link
to rotation (Wallerstein et al. 1994; De Medeiros et al. 2000).
 These works have shown a steady decline in lithium abundances from
spectral types F5III to F8III,
a wide
 spread around G0III and a gradual decline with temperature for stars
redward of
 such spectral type.
Wallerstein et al. (1994) have found that giant stars with \vsini
$>$\,50~\kms,
located in the (B$-$V)
 color interval from 0.40 to 0.70, present lithium abundance close
to the presumed primordial value, whereas slower
rotators show reduced lithium abundances in spite of their
 earlier spectral types. De Medeiros et al. (2000) have found a trend of
discontinuity in the
 distribution of the lithium abundances around the spectral type G0III,
paralleling the sudden
 decline observed in rotational velocity. The origin of such a
discontinuity is not yet well
 established but it seems to strongly depend on stellar mass (De Medeiros and
Mayor
1990).
 Nevertheless, the rotation-lithium connection in giant stars appears
to be a more complicated problem. In
 spite of the fact that high lithium content is associated with fast
rotation,
slow rotators present a large spread in the values of lithium abundance
(De Medeiros et al., 2000). In addition, this connection seems to depend on
the stellar mass, metallicity and age.  Another important question concerns
the level of dilution of lithium along the giant branch. While standard
theory predicts a factor of dilution of about 40 to 60 for 1\,\Msun and
2\,\Msun respectively (Iben, 1965a, 1965b),  the observations for these
stellar
masses  show that the factor of dilution of lithium is far higher,
reaching values as large as 400 to 1000. A more solid
 discussion about this factor of dilution requires more measurements of
lithium abundance for stars located near and at the Hertzsprung gap, in
particular for stars blueward of G0III, namely stars at the blue side of
the gap. The distribution of lithium abundances before stars evolve along
the giant branch may help us to understand the nature of the apparent
discontinuity in lithium content around G0III and to establish on a more
solid
basis the dilution factor of stars evolving along the giant branch.

In this work we present new measurements of lithium abundance for giant
stars in the spectral range F5III to G5III, typically stars located near
and along
the Hertzsprung gap. By combining these data with precise rotational
velocities we
analyse the rotation-lithium connection and the nature of the
discontinuities in
the distribution of these two stellar parameters.

\section{Observations}

For this study we have selected 54 luminosity class III giants with
spectral types
ranging from F5III to G5III. All these stars have been previously observed
with the CORAVEL spectrograph (Baranne et al., 1979) for rotational velocity
measurements and binarity signatures (De Medeiros and Mayor, 1999).

\subsection{Rotational velocity}

For most of the stars of the present sample, rotational velocity
measurements were
taken from
{\it The Catalog of Rotational and Radial Velocities for Evolved Stars} by De
Medeiros and
Mayor (1999). In this work, the authors present precise rotational velocity,
\vsini, for evolved
stars, obtained on the basis of observations acquired with the CORAVEL
spectrometers
(Baranne et al. 1979). For giant stars of luminosity class III in
particular, the
measured
rotational velocities  have an uncertainty of about
1.0\,\kms for stars with \vsini lower than about 30.0\,\kms. For faster
rotators,
the estimations
indicate an uncertainty of about 10\% on the measurements of \vsini. For a
few
stars of the sample, most
of them presenting evidence of high rotation, the \vsini value was determined
on the basis of the spectral synthesis carried out through this work.

\subsection{Spectroscopic observations}

The spectral region around the
lithium line at 6707.81\,\AA\, was observed with two different telescopes.
For
northern stars, high-resolution spectra  of the lithium region were
acquired with the AURELIE spectrograph (Gillet at al., 1994) mounted at the
1.52\,m telescope of the
Observatoire de Haute Provence (France). The spectrograph used a cooled
2048-photodiode
detector forming a 13~$\mu$m pixel linear array.
A grating with 1800 lines/mm was used, giving a mean dispersion of
4.7\,\AA/mm and  a resolving power around
45\,000 (at 6707\,\AA). For this instrumentation and the selected set-up, the
spectral
coverage was about 120\,\AA. The signal to noise ratio was always better
than 50.
For southern stars, high-resolution spectra were acquired with the Coud\'e
Echelle
Spectrometer (CES) in the long camera mode (Kapper \& Pasquini, 1996),
mounted at
the 1.44\,m CAT
telescope, at La Silla, ESO.
A RCA high resolution CCD with 640 x 1024 pixels was used as detector, with a
pixel size of
15 x 15~$\mu$m.
The dispersion was around
1.9\,\AA/mm and the resolving power was about 95\,000 (at 6707\,\AA). The
spectral
coverage for
this instrument was about 70~\AA\, and  the signal to noise ratio
was always better than 80.
For both observing runs, thorium lamps were observed before and after each
stellar
observation for
wavelength calibration, whereas 3 series of flat-field using an internal
lamp of
tungstene
were obtained during each night of observation. Flat-field corrections and
wavelength calibrations
were performed using the MIDAS package.

\section{Lithium abundances and spectrum synthesis}

In the present study, we adopted the spectral analysis method used
by L\`ebre at al. (1999) and Jasniewicz et al. (1999), in order to derive
the lithium abundances from the resonance LiI line
(6707.81\,\AA).
We refer to these authors for a
description of our abundance analysis assuming LTE. Nevertheless a few important
points are described below.

To compute the more appropriated synthetic spectra to fit
the observations, a first guess for the stellar parameters of our list
of giants has been estimated. Thus we have looked into the literature
and found effective temperatures (\Teff), gravities (\logg) and often
metallicites ([Fe/H]) for 37 stars (over 54), mainly
in Cayrel de Strobel et al. (2001), in Allende Prieto and Lambert (1999),
and in Alonso et al. (1999). For the rest of our sample (17 stars without
any published stellar parameters determination), we first estimated \Teff
 from (B-V) colour index (Flower, 1996),  and we have set \logg = 3.0, a
 microtuburlence velocity of 2\kms and
[Fe/H] = 0.0 which are values commonly adopted for Pop. I giant stars.

Synthetic spectra were then computed from these stellar
parameters and from the same line list and model atmospheres used
in  L\`ebre at al. (1999)
and Jasniewicz et al. (1999). The stellar parameters of all the giants
( mainly \Teff , [Fe/H] and \vsini) were then corrected,
when necessary, in order to improve the fit quality of the several
Fe~I and other metallic lines found in the observed spectral range.
The final adopted stellar parameters
and the derived Li abundances are presented in Table 1 for all the stars
of the
sample, except HD 156015 which was too cool to accurately derive its
effective
temperature with the adopted procedure (however, the absence of Li
signature is clear
for this star).
As already discussed by L\`ebre et al. (1999), the major source of
uncertainty
for this abundance analysis
is due to errors in the determination of the \Teff. We estimated that
effective temperature were derived with an uncertainty smaller
than $\pm$\,200~K. This leads to an error of less than 0.2~dex
on the derived metallicities and lithium abundances. We also checked
with the work of Carlsson et al. (1994) that non-LTE effects can be
neglected for the studied stars since these effects are always
much smaller than 0.1~dex.

\section{Results and Discussion}

\subsection{Rotation and \Ali~in the Hertzsprung gap}

The first step for the present analysis  was the construction of the HR
diagram to
locate the evolutionary stage of the stars composing the working sample.
In fact,
such a procedure
is important because in previous studies the only criterion for giant star
classification was the spectral type.
For this purpose, we have used HIPPARCOS (ESA 1997)
trigonometric parallax measurements and $V$ magnitudes  to compute
absolute magnitudes and
luminosities. Bolometric corrections $BC$ were determined from Flower (1996).
Evolutionary tracks were computed from the Toulouse-Geneva code for
stellar masses between 1 and 4\,\Msun and solar metallicity (see also
do Nascimento et al. 2000 for complementary informations).
The HR diagrams with the referred evolutionary tracks
are displayed in Figs. 1 and  2. In addition, these figures show
the behaviors of the lithium abundance and of the rotational velocity \vsini,
 respectively.
In these diagrams the dashed line indicates the evolutionary
region where the subgiant branch starts, corresponding to the hydrogen
exhaustion
in
stellar central regions, whereas  the dotted line represents the beginning
of the
ascent
along the red giant branch. Except for a few stars located along the subgiant
branch, the large
majority of stars in the present sample are effectively giants evolving
prior to
the ascent of
the red giant branch.

\begin{figure}
\centerline{\psfig{figure=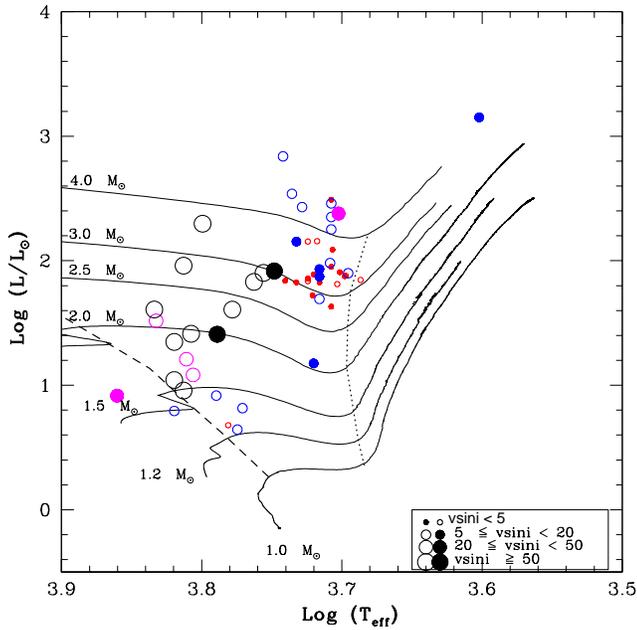,width=3.5truein,height=3.5truein}\hskip
0.1in}
\caption[]{Distribution of the rotational velocity of the stars in the HR
diagram.
Single and binary  stars are identified with  open and filled circles
respectively. The size of the circles is
proportional to the \vsini (in \kms).
Luminosities have been derived from the  Hipparcos parallaxes.
Evolutionary tracks
at [Fe/H]=0 are shown for stellar
masses between  1 and 4~M$_{\odot}$. The turnoff and the beginning of
the ascent on the red giant branch are indicated by the dashed and
dotted lines respectively.}
\label{HRrot}
\end{figure}

\begin{figure}
\centerline{\psfig{figure=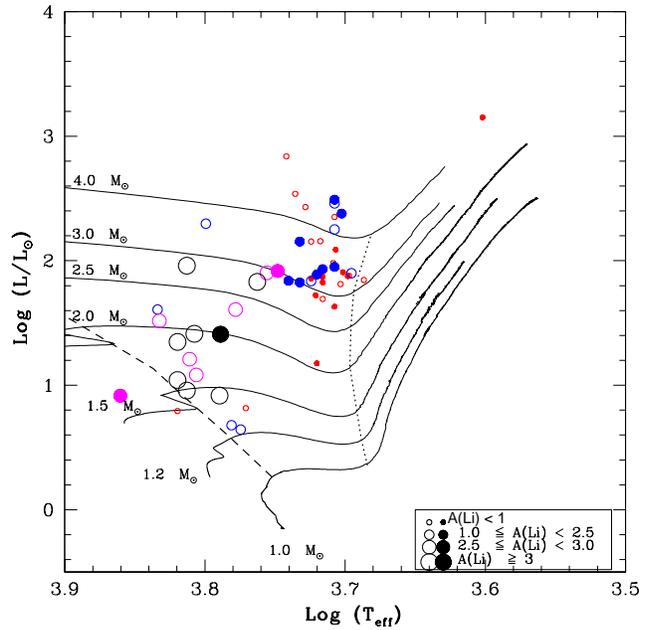,width=3.5truein,height=3.5truein}\hskip
0.1in}
\caption[]{Distribution of Li abundances in the HR diagram. Same as Fig.~1
except
that the symbol
size is proportional to the Li abundances.}
\label{HRlit}
\end{figure}

Figure~1 shows the well established rotational discontinuity for giants of
luminosity class III
(e.g. De Medeiros and Mayor 1990). As already shown by these authors, giants
blueward of the spectral
G0III (corresponding to the location of the rotational discontinuity, around
log(\Teff)\,$\approx$\,3.75), show a wide range of rotational velocity
from a few
\kms to
about one hundred times the solar rotation. Giants redward of such
spectral type
(G0III) are
essentially slow rotators, except for the synchronized binary systems and
a dozen
of late--G and K
single giants for which the origin of their high rotation is not yet fully
understood.
At least for stars with turnoff mass greater than 1.5\,\Msun, the sudden
decline
in rotation
appears simply to reflect the rapid increase of the moment of inertia as
the star
evolves across this
region of the HR diagram.

Figure~2 presents the behavior of lithium abundance, with a sudden decline in
\Ali~
near the same spectral region where the discontinuity in rotation is
observed.
Nevertheless,
it is quite clear from Figs.~1 and 2 that
the behavior of rotation and lithium content in the Hertzsprung gap, as
well as
the location of both
discontinuities,  strongly depends on stellar mass. Such a fact shows
that, for
giant stars, it
is not correct to define the location of the discontinuities in rotation
and in
\Ali~
at the same spectral type, as suggested by previous works based only on the
analysis of the
distributions of \Ali~ and \vsini versus spectral type or (B$-$V) color
index.
Single stars located on the blue side of the Hertzsprung gap show a large
spread
in the lithium
abundance, with values of \Ali~ ranging from less than 0.5 to 3.0~dex.
In contrast, stars located at the red side show essentially low values of
\Ali,
reflecting the dilution effects along this spectral region.

Clearly, one observes from Fig. 2 that stars with 2 to 3\,\Msun located
redward of
G0III
(log(\Teff)\,$\approx$\,3.75) show a factor of dilution of at least
600, which is far in excess  from the theoretical predictions. Such a fact
points
to an extra-mixing mechanism  occuring before the beginning of the ascent
of the
red giant branch.
The less massive giants show a similar disagreement between predicted and
observed
abundances.
Brown et al. (1989) have measured lithium abundances for a large sample of
late-G
to K giants,
with mean mass around 1.5 \,\Msun. For such stars, they found the mean
lithium
abundance lower than 0.0~dex, that is at least $\sim$~1.5\, dex below the
value
predicted by standard theory.
In Fig. 2 of the present paper, by taking into consideration the abundance
of lithium for stars with masses between 1.5\,\Msun and 3.0\,\Msun,
located at the
blue
side of the Hertzsprung gap, and referring to a mean value of \Ali~ lower
than
0.0~dex as estimated by
Brown et al. (1989), we observe that the factor of dilution for this
interval of
mass is even more
important than that observed for stars with 1.5\,\Msun.
\begin{figure}
\centerline{\psfig{figure=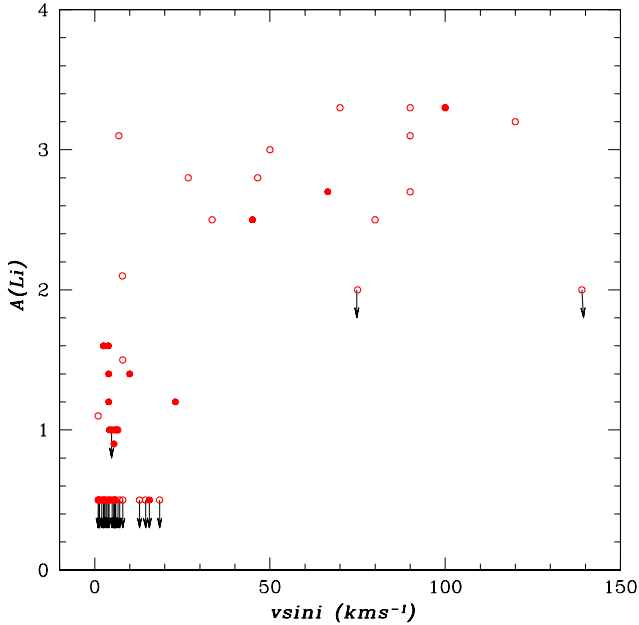,width=3.5truein,height=3.5truein}
\hskip 0.1in}
\vspace{1cm}
\caption[]{ Lithium abundance as a function of rotational velocity for giant
 stars in the Hertzsprung gap. Single and binary stars are identified by
open and filled circles respectively. Vertical arrows indicate \Ali~ upper
limits
}
\label{lirot}
\end{figure}

It is interesting to point out that the binary systems seem to follow the
same
tendency presented by
single stars. Redward of the spectral type G0III, the binary systems show
essentially low values
of \Ali, reflecting also the effects of dilution along the Hertzsprung gap.

\subsection{The connection lithium - rotation in the Hertzsprung gap}

The dependence of lithium content on rotation along the Hertzsprung gap
has been
already
reported in different works (Wallerstein et al., 1994; Alschuler 1975).
For the present
sample  of stars, this
trend is also clearly observed from Fig. 3, which displays  lithium
abundance as a
function of
rotational velocity. Stars with high rotation also present  high values of
\Ali.
In addition, the
present data show a trend also observed for other luminosity classes, as
well as
for other spectral
regions of giants. For slow rotators - in the present situation, for stars
with
\vsini lower
than about 20\,\kms - there is a large spread of the values of \Ali, with a
dispersion of at
least 3 magnitudes. Such a feature indicates that, also in the Hertzsprung
gap,
rotation
is not the unique parameter driving Li abundance. Let us recall that, in
spite of
the limited number
of binary systems with high \vsini value, the dependence of lithium on
rotation
seems to follow
the same trend observed for the single stars. We have also analyzed the
relation metallicity versus \vsini effects, looking for
possible effects of metallicity on the spread observed in the \Ali versus
rotation
relation, by using data presented in Table~1. No clear trend arises from
such an analysis.

\subsection{The connection lithium - deepening of the convective envelope}

Because the level of dilution of lithium depends on the level of convection
in the stars, we analyse here the behavior of lithium abundance as a function
of the deepening of the convective zone for our working sample. Such
an analysis sounds interesting because most of the stars are crossing the
Hertzsprung gap, where the convective zone is predicted to reach its most
important development. For our purposes, we have first estimated the mass of
each star $M_*$ from the HR diagram constructed for Figs.~1 and 2. Then,
we have
estimated the deepening in mass of the convective zone \Mcz, according to the
recipe applied by do Nascimento et al. (2000) and do Nascimento et al.
(2003).
In short, with the mass of each star in hand, we have placed such mass in the
convective zone mass deepening {$M_{\rm cz}/M_*$} versus Teff diagram
from do Nascimento et al. (2000) and estimated the parameter
{$M_{\rm cz}/M_*$} for each star. The individual values of $M_*$ and
{$M_{\rm cz}/M_*$} are listed in Table 1. The star HD 224342 is outside this
analysis because of its very large and uncertain luminosity and mass.
Figure 4 displays the deepening (in mass) of the convective envelope
versus effective temperature, for single and binary stars. In this
figure, the presence of the binary stars  should be analysed with caution due
to the large uncertainties in the estimation of mass for stars in binary
sytems using evolutionary tracks. However it is clear from this figure that
the sudden decline in \Ali~in the Hertzsprung gap is directly associated with
the rapid increase of the convective envelope in such region. Most of the
stars
with high Li content present an undeveloped convective zone, whereas stars
with low Li content show a developed convective zone.

\begin{figure}
\centerline{\psfig{figure=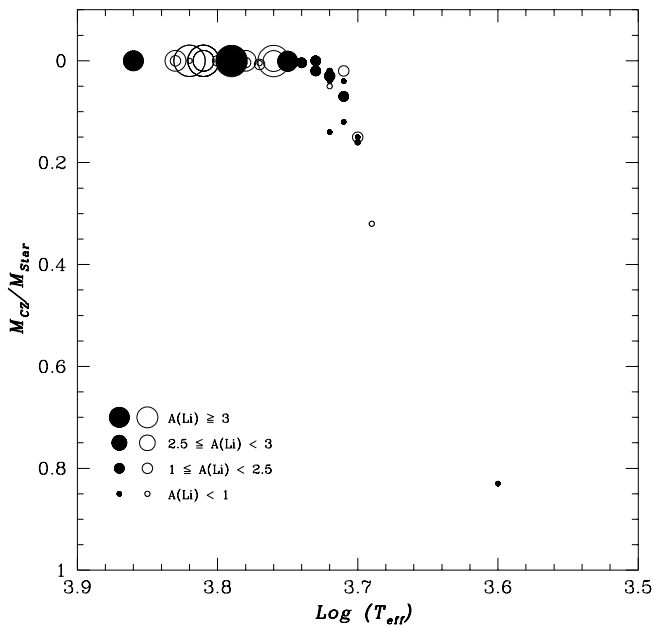,width=3.5truein,height=3.5truein}\hskip
0.1in}
\vspace{1cm}
\caption[]{The deepening (in mass) of the convective envelope as a
function of the
effective
temperature  for the stars in the present sample. Single and binary stars are
represented by open and filled circles respectively. The symbol size is
proportional
to the Li abundances quoted in the figure.}
\label{lizc}
\end{figure}

\section{Conclusions}

The present study brings precise abundances of Li for giant stars of
luminosity
class III
evolving across the  Hertzsprung gap. By combining these data with rotational
velocity and other
stellar parameters such as luminosity and effective temperature, we show a
trend for
the
existence of a sudden decline in \Ali~located around
log(\Teff)\,$\approx$\,3.75,
and
corresponding to the spectral type G0III, where the well established
rotational
discontinuity
for giants is defined.
Blueward of this region, most of the stars with mass greater than about
1.5\,\Msun
present a
trend for
high Li content, whereas for stars redward of this region Li is essentially
diluted.
The scenario of Li content in single stars seems to be followed by
binaries. The
dependence of
Li content on rotation, in the sense that fast rotators show high \Ali, is
also
confirmed by the present set of data. In addition, the large spread in
\Ali~for
slow
rotators observed for other luminosity classes and other spectral regions
occurs
also in the
Hertzsprung gap. Metallicity seems to have no effect on such a spread.
Finally, we have combined the \Ali~values with theoretical predictions on the
convective
zone deepening to analyse the extent of the level of convection on the Li
dilution.
It is clear from
this analysis that stars with high Li content have mostly undeveloped
convective
zones,
whereas stars with developed convective zone present low Li content.

\begin{table*}[f]
\begin{center}
\caption{The stars of the present working sample with their physical
parameters}
\label{tabela1}
\begin{tabular}{rlrcrrccrc}\hline\hline
HD    & (B-V) &$T_{\rm eff}$& log~$g$ &[Fe/H] &$v\sin i$ & M/M$_{\odot}$
&{$M_{\rm cz}/M_*$} & A$_{\rm Li}$
 & Rem  \\ \hline
87    & 0.901 & 5030     &  3.0   & -0.3  &  3.8     & 3.3 & 0.27 & $<$0.5
&  SB? \\
895   & 0.68  & 5100     &  3.0   & -0.3  &  2.5     & 2.8   & 0.12 &
$<$0.5 &  SB  \\
1671  & 0.442 & 6470     &  3.8   & -0.1  &  46.5    & 1.8 & 0.00 &    2.8
&      \\
6903  & 0.697 & 5700     &  3.0   & -0.2  & 90.0     & 2.9 & 0.00 &    2.7
& (1)   \\
17878 & 0.758 & 5300    &  2.7   &  0.1  &  2.6     & 3.8 & 0.02 & $<$0.5
&      \\
48737 & 0.443 & 6600     &  3.9   &  0.0  & 70.0     & 1.6 & 0.00 &    3.3
& (1)  \\
55052 & 0.397 & 6820     &  3.5   &  0.1  &  75.     & 2.2 & 0.00 & $<$2.0
&      \\
63208 & 0.571 & 5100     &  3.0   &  0.0  &  5.7     & 4.0 & 0.02 &    1.0
&      \\
65448 & 0.58  & 5250     &  3.0   &  0.0  &  2.5     & 3.3 & 0.02 &    1.6
& SB   \\
71369 & 0.856 & 5220     &  2.6   &  0.0  &  4.3     & 3.8 & 0.01 & $<$0.5
&      \\
72779 & 0.681 & 5790     &  3.0   &  0.0  & 90.0     & 2.7 & 0.00 &  3.3  
& (1)  \\
74485 & 0.935 & 4960     &  2.1   & -0.4  & 6.6      & 3.3 & 0.25 &   1.0 
&      \\
74874 & 0.685 & 5400     &  3.0   &  0.0  &  4.0     & 3.0 & 0.00 &  1.4  
& SBO \\
78715 & 0.89  & 5050     &  3.0   & -1.0  &  2.0     & 3.2 & 0.25 & $<$0.5
&     \\
81025 & 0.774 & 5200     &  3.0   & -1.0  &  5.0     & 2.9 & 0.02 & $<$0.5
&     \\
82210 & 0.781 & 5250     &  3.4   & -0.3  &  5.5     & 2.0 & 0.26 & 0.9   
& SBO \\
85945 & 0.895 & 5200     &  3.0   &  0.0  &  6.2     & 3.4 & 0.02 &  1.0  
&   SB  \\
92787 & 0.324 & 7250     &  4.2   & -0.3  & 45.0     & 1.6 & 0.00 &  2.5  
& (1,2),SB  \\
101133& 0.401 & 6800     &  3.4   &  0.0  & 33.5     & 2.1 & 0.00 &   2.5 
&      \\
107700& 0.515 & 5500     &  3.1   & -0.2  &  3.9     & 3.0 & 0.00 &   1.6 
&   SBO \\
108722& 0.445 & 6600     &  3.6   &  0.0  & 100.0    & 1.9 & 0.00 &   3.3 
&  (1) \\
111812& 0.681 & 5600     &  3.0   & -0.3  & 66.5     & 3.0 & 0.00 &   2.7 
&   SBO \\
119458& 0.857 & 5100     &  3.0   & -0.2  &  4.0     & 3.5 & 0.05 &   1.2 
&      SBO \\
121107& 0.845 & 5100     &  3.2   &  0.0  &  14.5    & 4.0 & 0.02 & $<$0.5
&      \\
136202& 0.54  & 6040     &  4.0   &  0.0  &  4.8     & 1.2 & 0.00 & $<$1.0
& \\
140438& 0.873 & 5110     &  3.0   &  0.1  &  5.8     & 3.5 & 0.05 & $<$0.5
& \\
151627& 0.877 & 5090     &  3.0   &  0.0  &  4.1     & 3.8 & 0.04 & $<$0.5
&   SB \\
152863& 0.921 & 4980     &  3.0   &  0.1  &  2.9     & 3.2 & 0.25 & $<$0.5
& \\
153751& 0.897 & 5040     &  2.0   & -0.3  &  23.     & 4.0 & 0.05 &   1.2 
&   SBO \\
155646& 0.504 & 6160     &  3.9   &  0.0  &  6.9     & 1.5 & 0.00 &  3.1  
& \\
156015& 1.164 & $<$4000     &        &       &  15.6    & 4.0 & 0.83 & 
$<$0.5  &   SBO \\
157358& 0.724 & 5300     &  3.0   &  0.0  &   1.0    & 3.1 & 0.05 &   1.1 
  & \\
159026& 0.51  & 6300     &  3.0   &  0.0  & 139.0    & 3.7 & 0.00 & 
$<$2.0  & \\
160365& 0.567 & 6150     &  3.4   &  0.0  & 100.0    & 2.0 & 0.00 &  3.3  
 & (1),SB \\
161239& 0.683 & 5900     &  3.8   &  0.4  &  5.9     & 1.4 & 0.00 & 
$<$0.5 & \\
169985& 0.487 & 5400     &  3.2   &  0.1  & 10.0     & 3.8 & 0.00 &   1.4 
 &   (1),SB \\
173920& 0.839 & 5350     &  3.0   &  0.0  &   8.0    & 4.0 & 0.01 & $<$0.5
 & \\
175492& 0.782 & 5100     &  3.0   &  0.0  &   4.2    & 4.0 & 0.01 &  1.0  
 &   SBO \\
182900& 0.456 & 6400     &  3.9   &  0.0  &   26.7   & 1.6 & 0.00 &   2.8 
 & (2) \\
185758& 0.777 & 5440     &  3.1   & -0.1  &  7.1     & 4.0 & 0.00 & 
$<$0.5 & \\
200039& 0.946 & 4990     &  3.0   & -0.4  &  1.0     & 3.1 & 0.25 & 
$<$0.5 &   SB? \\
200253& 0.994 & 5100     &  3.0   &  0.2  & 8.0      & 3.7 & 0.04 &  1.5  
 & (1)  \\
202447& 0.549 & 5300     &  3.2   & -0.4  & 1.3      & 3.1 & 0.03 & 
$<$0.5 & (2),SB \\
203574& 1.005 & 4860     &  3.0   & -0.4  & 1.0      & 3.0 & 0.35 & 
$<$0.5 &       \\
203842& 0.474 & 6420     &  3.0   & -0.3  & 90.0     & 2.0 & 0.00 &  3.1  
 & (1)  \\
208110& 0.786 & 5260     &  3.0   & -1.0  &   3.3     & 3.0 & 0.05 & 
$<$0.5 &   SB \\
209149& 0.463 & 6500     &  3.8   &  0.0  & 50.0      & 1.5 & 0.00 &  3.0 
  & (1)  \\
210459& 0.471 & 6500     &  3.0   &  0.0  & 120.0     & 2.8 & 0.00 &  3.2 
  & (1)  \\
214558& 0.771 & 5200     &  3.0   & -0.4  &   1.4     & 3.1 & 0.03 & 
$<$0.5 &   SB \\
215648& 0.502 & 5950     &  4.0   & -0.4  &   7.9     & 1.2 & 0.00 &  2.1 
  & \\
218658& 0.802 & 5200     &  2.9   & -0.2  &   5.5     & 3.3 & 0.02 & 
$<$0.5 & SBO \\
220657& 0.617 & 6000     &  3.0   &  0.0  & 80.0      & 2.4 & 0.00 &  2.5 
  & (1)  \\
223346& 0.461 & 6600     &  4.1   &  0.0  &   18.5    & 1.5 & 0.00 & 
$<$0.5 &    \\
224342& 0.712 & 5520     &  2.0   & -1.0  &   12.8    & 4.5 & $--$ & 
$<$0.5 &    \\

\hline
\end{tabular}
\end{center}
 (1) \vsini  determined from the spectral synthesis.
 (2) Uncertain \Teff or suspected binary
\end{table*}

\begin{acknowledgements}
JRM warmly acknowledges colleagues
of the Cote d'Azur Observatory at Nice, where a large part of this work was
prepared during his stay in France.
This work has been supported by continuous grants from
the CNPq Brazilian Agency and by funds from the french Programme National de
Physique Stellaire
(INSU/CNRS). J.D.N.Jr. acknowledges the CNPq for the fellowship PROFIX
540461/01-6.
We also gratefully acknowledge C. Ferrari for her wonderful software {\it
tiramisu}.
\end{acknowledgements}

\end{document}